\documentclass[aps,prl,twocolumn,superscriptaddress]{revtex4-1}
\usepackage{amssymb}    % need for subequations
\usepackage{amsmath} 
\usepackage{graphicx}   % need for figures
\usepackage{verbatim}   % useful for program listings
\usepackage{color}      % use if color is used in text
\usepackage{subfigure}  % use for side-by-side figures
\usepackage{hyperref}   % use for hypertext links, including those to external documents and URLs
\usepackage{epsfig}
\usepackage[utf8]{inputenc}
\begin{document}

%\preprint{}

\title{Formation of Vacancies in Si- and Ge-based Clathrates:\\ Role of Electron Localization and Symmetry Breaking}


\author{Amrita Bhattacharya}
\affiliation{Fritz Haber Institute of the Max Planck Society, Faradayweg 4-6, 14195 Berlin, Germany}
\author{Christian Carbogno}
\affiliation{Fritz Haber Institute of the Max Planck Society, Faradayweg 4-6, 14195 Berlin, Germany}
\author{Bodo B$\ddot{\mathrm{o}}$hme}
\affiliation{Max Planck Insititute for Chemical Physics of Solids, N$\ddot{o}$thnitzer Str.
40, 01187, Dresden, Germany}
\author{Michael Baitinger}
\affiliation{Max Planck Insititute for Chemical Physics of Solids, N$\ddot{o}$thnitzer Str.
40, 01187, Dresden, Germany}
\author{Yuri Grin}
\affiliation{Max Planck Insititute for Chemical Physics of Solids, N$\ddot{o}$thnitzer Str.
40, 01187, Dresden, Germany}
\author{Matthias Scheffler}
\affiliation{Fritz Haber Institute of the Max Planck Society, Faradayweg 4-6, 14195 Berlin, Germany}
\affiliation{Department of Chemistry and Biochemistry, University of California at Santa Barbara, CA 93106, USA}
\affiliation{Materials Department, University of California at Santa Barbara, CA 93106, USA}
%\email[]{amrita@fhi-berlin.mpg.de}
%\thanks{}
%\affiliation{Fritz Haber Institut of the Max Planck Gesellschaft}


\date{\today}

\begin{abstract}
The formation of framework vacancies in Si- and Ge-based type-I clathrates is studied as function of filling the cages with K and Ba atoms using density-functional theory. Our analysis reveals the relevance of structural disorder, geometric relaxation, electronic saturation, as well as vibrational and configurational entropy. In the Si clathrates we find that vacancies are unstable, but very differently, in Ge clathrates up to three vacancies per unit cell can be stabilized. This contrasting 
behavior is largely driven by the different energy gain on populating the electronic vacancy states, which originates from the different degree of  
localization of the valence orbitals of Si and Ge. This also actuates a qualitatively different atomic relaxation of the framework.
\end{abstract}

% insert suggested PACS numbers in braces on next line \pacs{}
%\keywords{}

\maketitle

\begin{figure}
\includegraphics[width=0.9\linewidth]{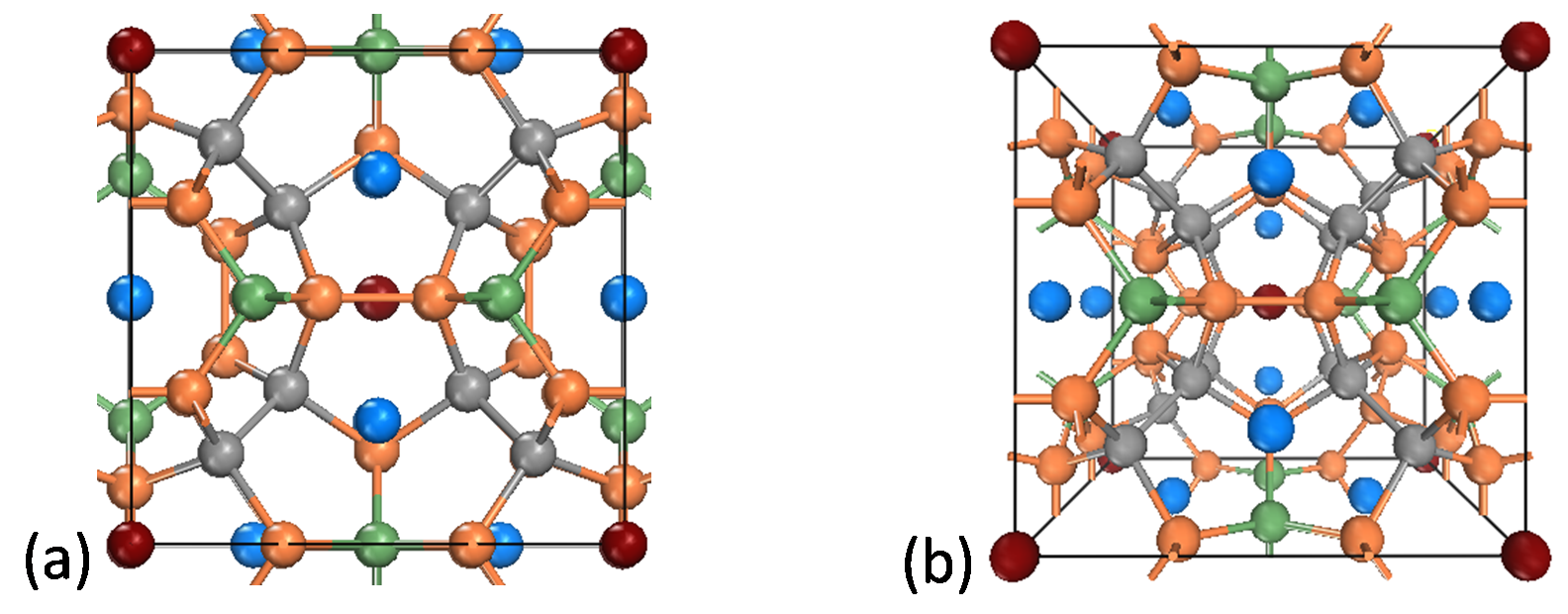}
\caption{Crystal structure~(space group~$Pm\bar{3}n$) of type-I clathrates~\cite{Nolas2014}. Colors denote the Wyckoff sites for the host~($6c$~green, $16i$~grey, and $24k$~orange) 
and the guest atoms~($2a$ red, $6d$ blue).}
\label{FigStruct}
\end{figure}

Clathrates are compounds with complex and large cage-like crystal structures~({\it hosts})
that can be filled with {\it guest} atoms or molecules~(Fig.~\ref{FigStruct})~\cite{Takabatake2014}.
Charge and heat transport in intermetallic clathrates has been studied intensively over the last decade~\cite{Nolas2014},
since filling would allow to increase their thermoelectric efficiency~\cite{Christensen2010}. The uttermost majority of filled clathrates 
are, however, metallic~\cite{Hagenmuller1970,Ramachandran2000,VonSchnering2011,Yamanaka2000,Lortz2008,Castillo2015} and thus unsuitable for this application in their pristine form. 
Nonetheless, puzzling exceptions exist,~e.g.,~Ge$_{46}$ filled with K:
After decades of experiments~\cite{Hagenmuller1970,Ramachandran2000,VonSchnering2011},
its semiconducting character was recently explained by a careful structure analysis~\cite{Beekman2007},
which revealed  a high vacancy concentration~($\sim 4$\%).
For this case, the puzzle appears to be resolved~\cite{Schnering1985,Schnering1988}.
However, high-quality synthesis and experimental analysis have remained challenging in this field.
In particular, no fundamental understanding of the mechanism that determines  
vacancy formation and thus  composition, structure, and electronic character upon filling exists. 
Therefore, these properties of clathrates are still unpredictable in practice.

The most prominent example are the isoelectronic Si and Ge type-I clathrates~(see Fig.~\ref{FigStruct}):
These clathrates exhibit comparable properties in the unfilled case~(cf.~Tab.~\ref{Par} and Ref.~\cite{Connetable2010}),
but behave remarkably different upon filling,~e.g.,~with K or Ba guests. In the case of Si, the framework remains intact and
metallic behavior~(e.g.~K$_8$Si$_{46}$~\cite{Hagenmuller1970,Ramachandran2000,VonSchnering2011} and Ba$_{8}$Si$_{46}$~\cite{Yamanaka2000,Lortz2008,Castillo2015}) results. 
In the case of Ge, vacancies~$\square$ occur in the host framework: 
In line with Zintl's concept, a heavily doped semiconductor~K$_8$Ge$_{44}\square_2$ featuring 
two framework vacancies was found experimentally for monovalent K guests~\cite{Beekman2007,VonSchnering2011}.
For divalent Ba guests, however, experiments~\cite{Cabrera2004,Aydemir2010} found a metallic Ba$_8$Ge$_{43}\square_3$ compound with three 
vacancies, but not four, as expected from the increased Ba valence.
In this work, we present a quantitatively reliable theoretical prediction of the structure, vacancy concentration~(composition),
electronic character, and thermodynamic stability for these binary K/Ba-Si/Ge systems. These findings explain the existing 
experimental results and provide an atomistic mechanism for the contrasting behavior of Si and Ge clathrates.

Density-functional theory~(DFT)~\cite{Hohenberg1964,Kohn1965} calculations were performed with {\it FHI-aims}~\cite{Blum2009,Knuth2015}, an all-electron, 
full-potential electronic-structure code that uses numeric, atom-centered basis sets. Numerical 
settings 
were chosen to achieve a convergence in energy differences better than $10^{-3}$ eV/atom~(see Suppl. Mat.).
To ensure that our findings are independent from the chosen treatment for exchange and correlation~(xc), we 
compare how various xc-functionals describe both the equilibrium properties of the empty~$F_{46}$ clathrates~($F=$~Si, Ge) and the charged vacancy~\cite{Weinert1987}. 
Its formation energy is calculated using the total energy difference~\cite{Freysoldt2014}
\begin{equation}
\label{E-v}
E_\square^q = E(F_{45}\square_1^q) -E(F_{46})+ \frac{E(F_2^{\mathrm{Dia}})}{2} + q(\mu_e - \mathrm{VBM})\;
\end{equation}
between the defective framework~$E(F_{45}\square_1^q)$ with charge~$q$ and the pristine clathrate~$E(F_{46})$.
The diamond phase $E(F_2^{\mathrm{Dia}})$ and the electron chemical potential~$\mu_e$ relative to the 
valence band maximum~VBM act as thermodynamic reservoirs. 
The VBM of the defective and of the pristine clathrate~($\mathrm{VBM}^\mathrm{{Pr}}$) are referenced using the core level shift $\Delta V$ between 
$F_{45}\square_1^q$ and $F_{46}$: $\mathrm{VBM} = \mathrm{VBM}^\mathrm{{Pr}} + \Delta V$. 
Charge transition levels $I_{q/q'}$, which quantify the energy involved in charging the vacancy~\cite{Freysoldt2014}, are computed 
using the value of $\mu_e$ at which $F_{45}\square_1^{q'}$ and $F_{45}\square_1^q$ are in equilibrium 
\begin{equation}
I_{q/q'} = \frac{E(F_{45}\square_1^{q'})- E(F_{45}\square_1^{q})}{q-q'} + \mathrm{VBM}\;.
\end{equation}
\begin{table}[t]
\begin{ruledtabular}
\begin{tabular}{lccccc|ccccc}
          & \multicolumn{5}{c|}{Si$_{46}$}  &  \multicolumn{5}{c}{Ge$_{46}$}\\
          & $a_0$    &  $E_\mathrm{c}$ & $\epsilon_g$ & $E_{\square}^0$ & $I_{0/4-}$ & $a_0$ & $E_\mathrm{c}$  &  $\epsilon_g$ &  $E_{\square}^0 $  & $I_{0/4-}$ \\ \hline      
LDA       &    10.11 &   3.8          & 1.14  &   3.33    & 0.93    & 10.49 & 2.2            &  1.25  &    2.97      & 0.60    \\
PBEsol    &    10.17 &   3.4          & 1.21  &   3.35    & 0.95    & 10.59 & 1.9            &  1.23  &    2.92      & 0.58    \\
PBE       &    10.23 &   2.7          & 1.33  &   3.43    & 1.10    & 10.74 & 1.5            &  1.12  &    2.80      & 0.57    \\
RPBE      &    10.30 &   2.0          & 1.43  &   3.50    & 1.08    & 10.84 & 1.0            &  1.09  &    2.77      & 0.58    \\\hline
HSE06     &    10.18 &   3.4          & 1.90  &   3.85    & 1.48    & 10.62 & 1.7            &  1.89  &    3.32      & 0.95    \\
\end{tabular}
\end{ruledtabular}
\caption{\label{Par} Properties of empty Si$_{46}$ and Ge$_{46}$ clathrates computed in the 46 atom unit cell for various xc-functional: Lattice parameter~$a_0$~($\mbox{\AA}$), cohesive energy $E_\mathrm{c}$~(eV), Kohn-Sham band gap~$\epsilon_g$~(eV), neutral vacancy formation energy~$E_{\square}^0$~(eV), and charge transition level~$I_{0/4-}$ with respect to the top of the valence band. The atomic chemical potential~(for $E_c$ and $E_\square^0$) is that of the diamond structure.}
\end{table}

For the local-density approximation~(LDA)~\cite{Perdew1992}), variants 
of the generalized gradient approximation~(PBE~\cite{Perdew1996}, PBEsol~\cite{Perdew2008}, RPBE~\cite{Hammer1999}),
and the computationally more involved HSE06 functional~\cite{Krukau2006} that incorporates
a fraction of exact exchange, Tab.~\ref{Par} lists
the lattice parameter~$a_0$, cohesive energy~$E_\mathrm{c}$, and band gap~$\epsilon_g$ for the empty $F_{46}$
clathrates as well as the formation energy for a neutral vacancy~$E_{\square}^0$ at the 6$c$~site
and its transition level~$I_{0/4-}$, which characterizes the typical charge state in these clathrates~(see below).
We find the typical over/underbinding of LDA/PBE for $a_0$ and~$E_\mathrm{c}$,
whereby PBEsol gives results similar to HSE06~(differences are $<0.5$\% for $a_0$ and $<5$~meV/atom for $E_c$). 
Deviations in the vacancy formation energies~$E_{\square}^0$ and transition levels~$I_{0/4-}$ are, however, noticeable. 
In the following, PBEsol was thus employed in the structural search. We explicitly checked that the relative energetic 
ordering of the thereby identified compositions is retained with HSE06~(see Suppl.~Mat.).

\begin{figure}
\includegraphics[width=0.95\linewidth]{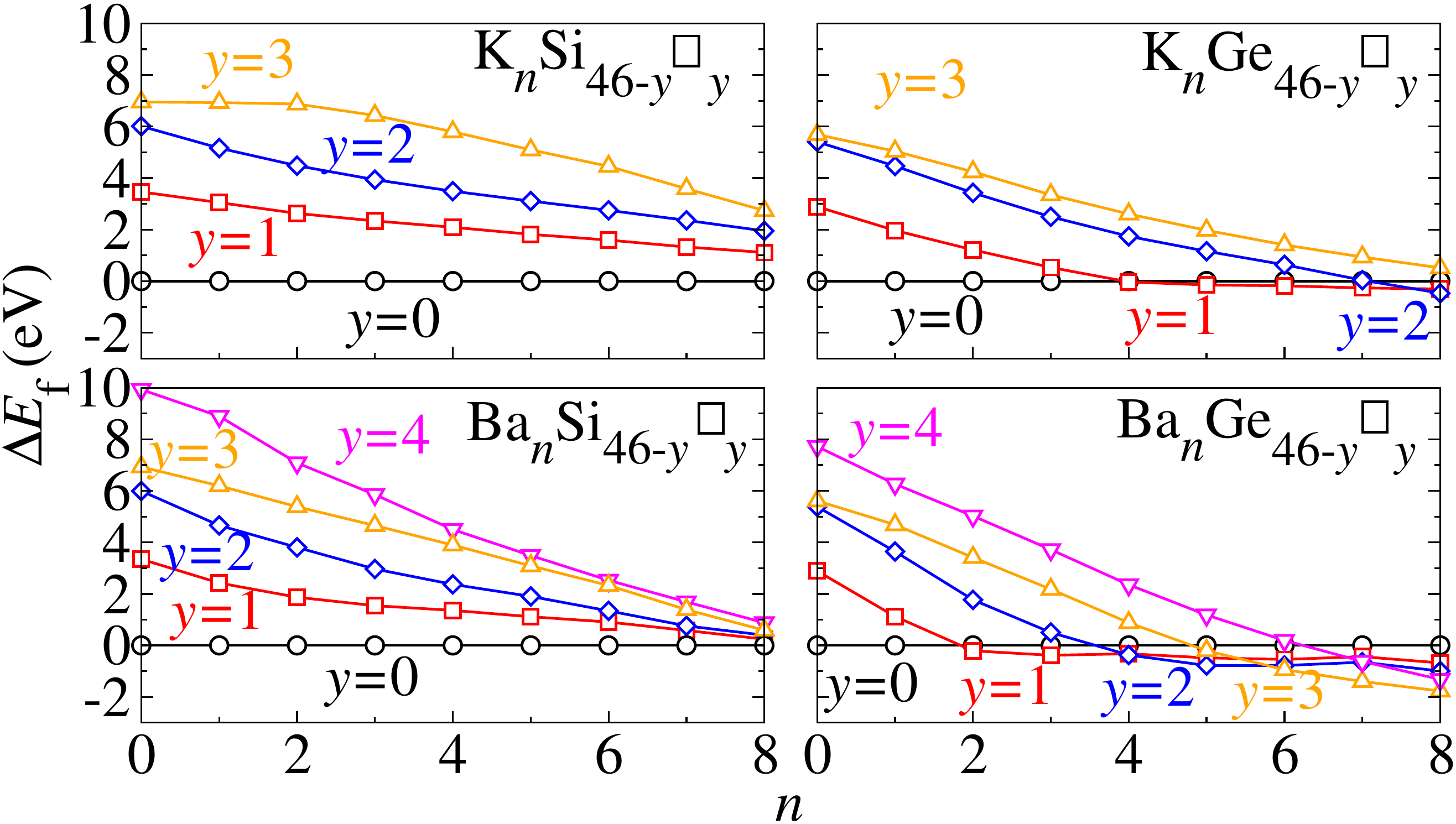}
\caption{\label{EF} Calculated~(DFT-PBEsol) formation-energy difference $\Delta E_\mathrm{f} =  E_\mathrm{f}(G_n F_{46-y}\square_y)- E_\mathrm{f}(G_nF_{46})$ as function of the filling~$n$ with K and Ba guests.}
\end{figure}

To determine the stable vacancy concentrations in Si and Ge frameworks filled with K or Ba guests~($G$), we have first identified the energetically favorable configurations for all compositions~$G_nF_{46-y}\square_y$ with $n \in [0,8]$~guests and $y \in [0,4]$~vacancies using
an iterative strategy that required $\sim$ 1400 full structural relaxations: Starting from the completely filled clathrate,
we have first identified the most favorable vacancy sites by scanning over all possible framework positions. 
Then, we stepwise removed guests from the compositions with fully occupied and defective framework, again scanning
over all available guest sites. For any subsequent composition, we retained the already identified guest and vacancy sites and thus limited 
the scanning to the remaining available positions. Eventually, we computed the formation energies of these compositions
using the stochiometrically balanced energy difference 
\begin{eqnarray}
\label{E_form}
E_\mathrm{f}(n,y) & = & E(G_nF_{46-y}\square_y)\\
&&-\frac{46-y-n \cdot x}{2} \cdot E(F_2^{\mathrm{Dia}})-  n\cdot E(G_1F_x)\nonumber
\end{eqnarray}
between filled and/or defective clathrate~$E(G_nF_{46-y}\square_y)$
and the reservoirs for framework~$F$ and guest atoms~$G$,~i.e.,~$F_2^{\mathrm{Dia}}$ and the thermodynamically stable neighboring 
phases~$G_1F_x$~(K$_4$Si$_4$~\cite{Schnering2005}, BaSi$_2$~\cite{Goebel2009}, K$_4$Ge$_4$~\cite{VonSchnering2005}, and Ba$_6$Ge$_{25}$~\cite{Carrillo2000}).
As the formation energy differences in Fig.~\ref{EF} show, vacancy formation is always energetically unfavorable in Si clathrates by $\Delta E_\mathrm{f}>0.3$~eV/vacancy, but
not in Ge clathrates: In the fully filled case, di-vacancy~(K$_8$Ge$_{44}\square_2$) and tri-vacancy formation~(Ba$_8$Ge$_{43}\square_3$) 
are energetically favorable by $\Delta E_\mathrm{f}< -0.1$~eV/vacancy; partial filling~($n<8$) makes smaller vacancy concentrations preferable~\footnote{In all cases, the first two vacancies are most favorable at $6c$, the third and fourth at $24k$ positions. However, taking into account superstructure formation in Ba$_8$Ge$_{43}\square_3$~\cite{Aydemir2010}, the experimentally reported vacancy configuration~($2\times 2\times 2$ superstructure; space group~$Ia\bar{3}d$) is found to be slightly more stable by $0.05$~(PBEsol) and $0.07$~(HSE06)~eV/f.u..}.

\begin{figure}[t]
\includegraphics[width=0.95\linewidth]{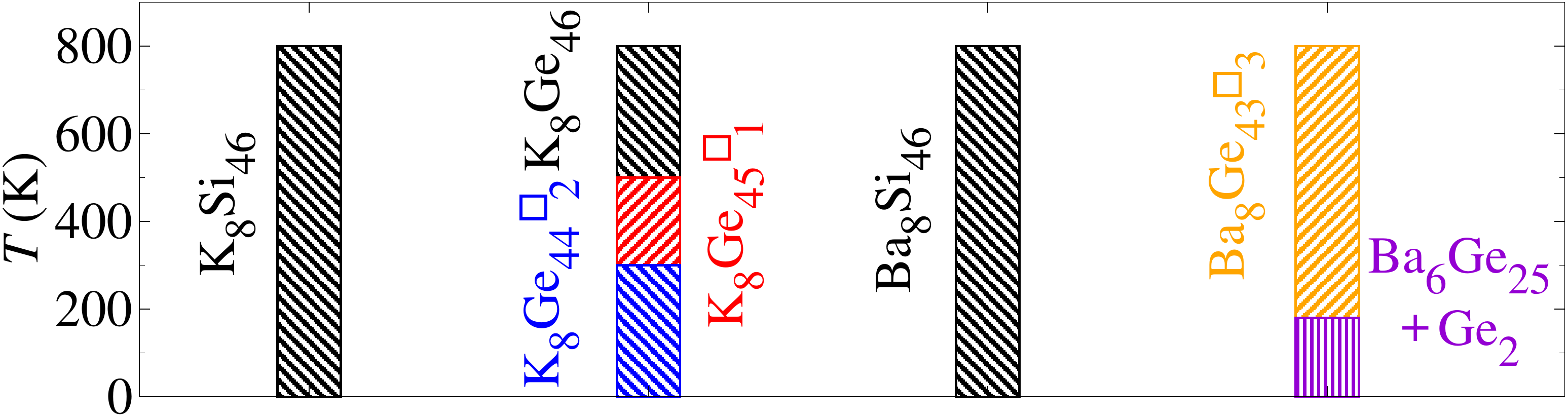}
\caption{\label{phase_T} Most favorable compositions of K/Ba-filled Si/Ge clathrates as a function of temperature~$T$ computed
at the DFT-PBEsol level.}
\end{figure}

Fig.~\ref{EF} also reveals that vacancy formation energies generally decrease with increasing filling, so that the question arises, if
the fully filled clathrates identified as stable at 0~K are also the most stable ones at finite temperatures. To clarify, 
we have computed their thermodynamic stability by accounting for configuration entropy and vibrational free energies~(see Suppl. Mat.). As shown in Fig.~\ref{phase_T}, we find that the compositions discussed above~(K$_8$Si$_{46}$, K$_8$Ge$_{44}\square_2$, Ba$_8$Si$_{46}$, Ba$_8$Ge$_{43}\square_3$) are also stable at room temperature. Interestingly, we find that vacancies in the K-filled Ge clathrate become less favorable with increasing temperature, which reflects that their formation is energetically favorable but entropically adverse 
due to the harder vibrations present in defective frameworks~(see Suppl. Mat.). Since this compound is typically synthesized at temperatures $>600$~K~\cite{Beekman2007,Hagenmuller1970,Ramachandran2000,VonSchnering2011}, this explains reports of less than two vacancies per unit cell~\cite{Llanos1984}. Accordingly, our calculations correctly reproduce the trends in formation and phase stability for all compositions observed experimentally~\cite{Ramachandran2000,Yamanaka2000,Beekman2007}, including the eutectoid decomposition of Ba$_8$Ge$_{43}\square_3$~\cite{Aydemir2010}.

Furthermore, our calculations consistently reproduce the electronic character found for these compositions experimentally: The guests donate electrons 
to the framework, which leads to a partial filling of the conduction band and thus to a metallic electronic structure in the stable, vacancy-free Si clathrates~(8 and 16 charge carriers/f.u.~for K and Ba with $n=8$). Conversely, 
the vacancies in the Ge framework accommodate up to four surplus electrons each, as suggested by qualitative models~\cite{Schnering1985,Schnering1988}. Due to the completely occupied vacancy states, K$_8$Ge$_{44}\square_2$ is 
a semiconductor with a band gap of 0.18/0.28 eV (PBEsol/HSE06). Similarly, all vacancy states are occupied in Ba$_8$Ge$_{43}\square_3$, but the four additional electrons accommodated in the conduction band lead to a metallic electronic structure.

Our studies confirm that filled Si- and Ge-based clathrates behave remarkably even qualitatively differently in spite of their isoelectronicity. 
Fig.~\ref{EF} also suggests the mechanism that stabilizes vacancies in Ge, but not in Si:
In Ge, the slope $\partial \Delta E_\mathrm{f}(n,y)/\partial n$ changes significantly whenever the number of guest atoms $n$ matches $n=4y/z$~($z$ is the valency of the guest cation),~e.g.,~for K$_4$Ge$_{45}\square_1$ or Ba$_4$Ge$_{44}\square_2$. 
Regardless of the guests' species, the Ge compositions meeting this condition are energetically favorable and are the only ones that exhibit semiconducting 
 character. Conversely, such distinct changes in the slope are not observed in Si, for which $\partial \Delta E_\mathrm{f}(n,y)/\partial n$ 
is almost independent on the number of guests. Since $\Delta E_\mathrm{f}(n,y)$ is defined as the energy difference between the defective 
and the completely occupied host framework, its slope~$\partial \Delta E_\mathrm{f}(n,y)/\partial n$ is related to the energy gain stemming from
 charging the vacancy.  

\begin{figure}[t]
\includegraphics[width=0.95\linewidth]{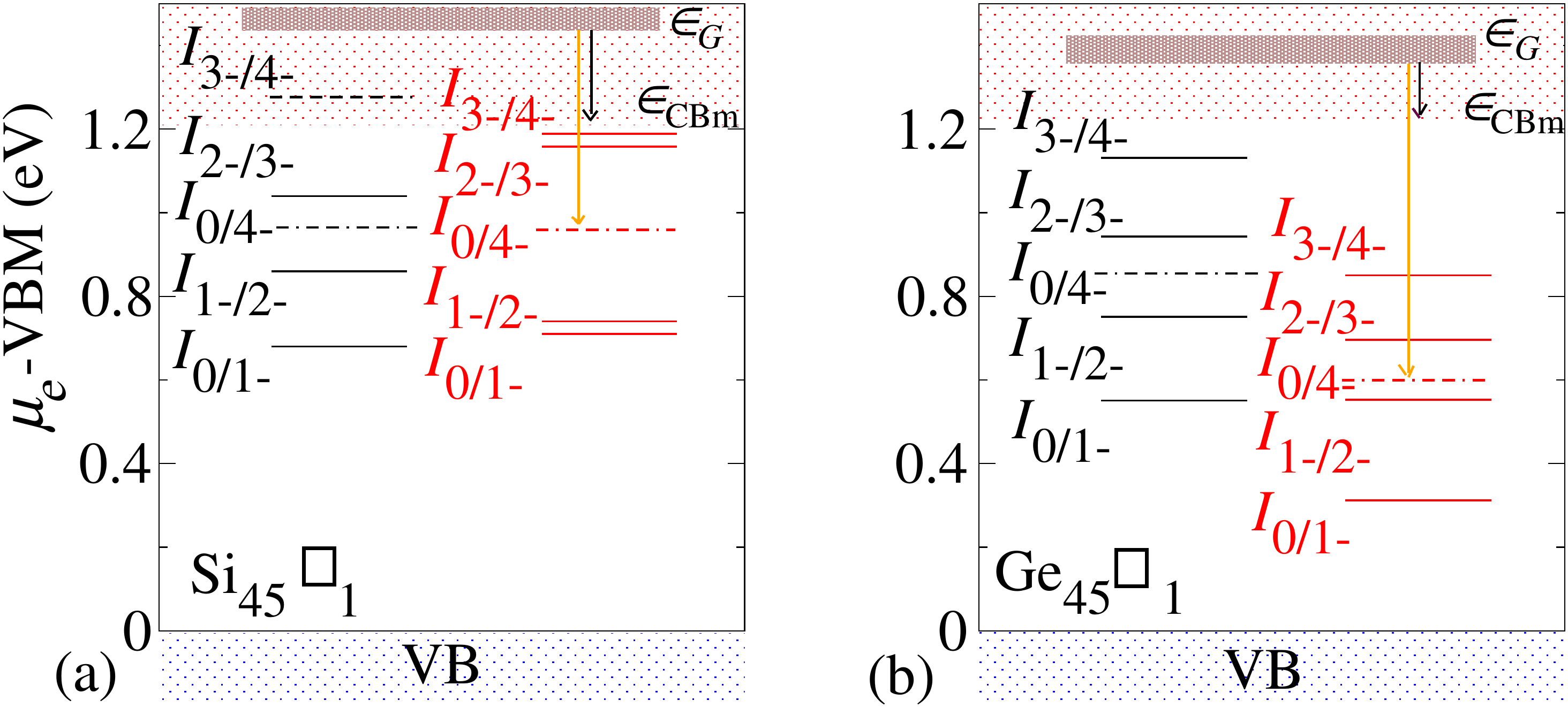}
\caption{\label{ion_level} Calculated~(DFT-PBEsol) charge transition levels $I_{q/q'}$ of (a)~$\mathrm{Si}_{45}\square_1$ and (b)~$\mathrm{Ge}_{45}\square_1$ with the unrelaxed~(black) and fully relaxed~(red) geometry. Schematically, the positions of the guest levels~$\epsilon_G$ in the conduction band are drawn as well to highlight the energy gains $E_\mathrm{CBm}$ and $E_\square$ associated with electron transfer from $\epsilon_G$ to the conduction band minimum~$\epsilon_\mathrm{CBm}$~(black arrow) and to the vacancy's ionization level~$I_{0/4-}$~(orange arrow).}
\end{figure}

In more detail, this mechanism can be rationalized by inspecting the electronic structure of these defected clathrates:
As sketched in Fig.~\ref{ion_level}, the guests lose their valence electrons in state~$\epsilon_G$ 
to the energetically lower lying conduction band minimum~$\epsilon_\mathrm{CBm}$ or -- if available -- to the vacancy states~$I_{q/q'}$ in the band gap.
The respective energy gains are $E_\mathrm{CBm} =\epsilon_G-\epsilon_\mathrm{CBm}$ and $E_{q/q'} = \epsilon_G-I_{q/q'}$. In first order approximation, their 
difference $\Delta E_{q/q'} =E_\mathrm{CBm}-E_{q/q'}$ determines the slope $\partial\Delta E_\mathrm{f}(n,y)/\partial n$ for~$n \leqslant 4y/z$. 
Please note that by definition $\Delta E_{q/q'}$ is not particularly sensitive on the guests' species~(see Suppl.~Mat.),
but is predominantly determined by the host. Surprisingly, this reveals that the contrasting tendency to suppress/form vacancies 
in Si and Ge frameworks is largely controlled by the charge transition levels of their vacancies. Indeed, Fig.~\ref{EF} and~\ref{ion_level} consistently show that 
both $\partial\Delta E_\mathrm{f}(n,y)/\partial n$ and $\Delta E_{q/q'}$ are much smaller in Si than in Ge. Accordingly, 
the different properties of a single vacancy  in guest-free Si$_{45}\square_1$ and Ge$_{45}\square_1$ also allow
to rationalize the underlying mechanism:
Quantitatively, the energetic cost~$E_\square^0$ to create a neutral vacancy is indeed slightly higher by $\approx$~0.4~eV in Si~(cf.~Tab.~\ref{Par}); 
the energetic gain~$\Delta E_{0/4-} = -4(\epsilon_\mathrm{CBm}-I_{0/4-})$ obtained from fully charging the vacancy  
is, however, distinctly larger in Ge by $\approx$ -1.2 eV and thus exceeds the cost of creating a vacancy -- but only if geometric relaxations are accounted for~(cf. 
Fig.~\ref{ion_level}). In the unrelaxed case, 
the charge transition levels~$I_{q/q'}$ show only slight quantitative differences 
for Si and Ge, but the exact same qualitative behavior. Upon geometry relaxation, however, 
the charge transition levels in Si form energetically almost degenerate, zero-$\mathrm{U}$~\footnote{$U$ is calculated from 
Eq.~\ref{E-v} using the expression $U = E_\square^{q+2}(\mu_e=0) + E_\square^{q}(\mu_e=0)-2 E_\square^{q+1}(\mu_e=0)$} 
pairs~$I_{0/1-}$ and $I_{1-/2-}$ as well as $I_{2-/3-}$ and $I_{3-/4-}$, whereby $I_{0/4-}$ is virtually unaltered.
Conversely, the individual levels including $I_{0/4-}$ undergo distinct shifts to lower 
energies in Ge, which explains the large overall energy gain~$\Delta E_{q/q'}$. 
This demonstrates that the contrasting behavior of Si and Ge clathrates is neither driven by  guest-host interactions 
nor by the different cost of creating neutral vacancies, but is predominantly determined by electron transfer processes 
upon charging.

\begin{figure}
\includegraphics[width=0.9\linewidth]{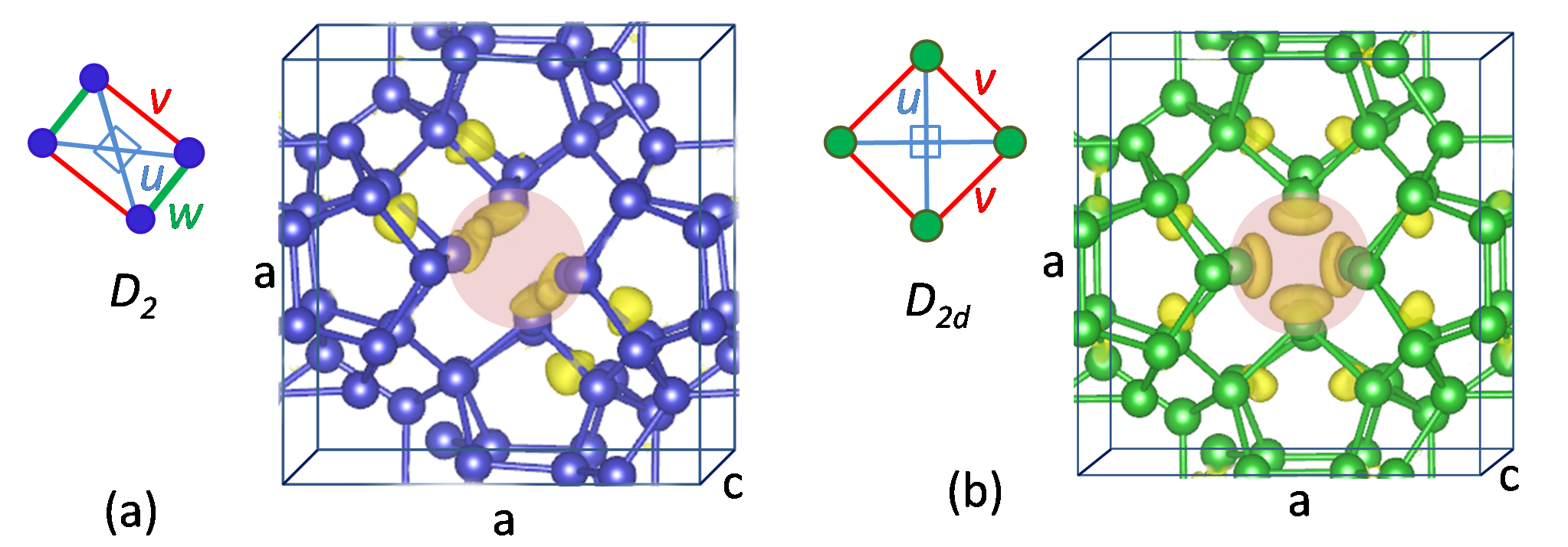}
\caption{\label{Iso} Calculated~(DFT-PBEsol) geometries of the relaxed $F_{45}\square_1^{4-}$ vacancy in empty (a)~Si and (b)~Ge clathrates. 
The isosurface~(isovalue~$0.015$~e$\mbox{\AA}^{-3}$) of the charge density 
difference with respect to the uncharged system~$F_{45}\square_1^0$ is shown
and the vacancy region is highlighted.} 
\end{figure}

These electron transfer processes result from the different character of the Si and Ge transition levels, which arise 
from different geometric distortions. Qualitatively, this is reflected in the electron distribution around the vacancy~(Fig.~\ref{Iso}). 
Quantitatively, we use the interatomic distances $u,~v,~\mathrm{and}~w$ between the vacancys' neighbors~(cf.~Fig.~\ref{Iso} and~\cite{Baraff1980,Fazzio2000}) and the lattice parameter ratio~$c/a$ to characterize the local and global symmetry. 
In the unrelaxed case~($c/a=1$), the neutral vacancy exhibits a global $D_{2d}$ symmetry with nearest neighbor distances $u$~(3.90~\mbox{\AA} in Si, 4.07~\mbox{\AA} in Ge) and $v=w$~($3.86~\mbox{\AA}$ in Si, $4.02~\mbox{\AA}$ in Ge). In the case of Ge$_{45}\square_1$, the local $D_{2d}$ symmetry~($u=3.15~\mbox{\AA}$, $v=w=3.39~\mbox{\AA}$) and the cubic lattice~($c/a\le1.01$) are essentially retained when relaxing the fully charged vacancy.
Conversely, we find that the fully charged Si vacancy features a local $D_2$~symmetry~($u=2.98~\mbox{\AA}$, $v=3.55~\mbox{\AA}$) with a drastically 
shortened pair of distances~$w=3.07~\mbox{\AA}$. Also,  a {\bf global} break in lattice symmetry (tilting~$\gamma = 92^{\circ}$ and~$c/a=1.02$) occurs, which results 
in a splitting of the charge density around the vacancy in two distinct lobes. 
The same qualitative behavior also occurs in filled clathrates (see Suppl.~Mat.).
Please note that contrasting local relaxation patterns have been reported for the vacancy in the respective Si and Ge diamond phases~\cite{Baraff1980,Fazzio2000}, too, but not a global symmetry break as in this case.

\begin{figure}
\includegraphics[width=0.95\linewidth]{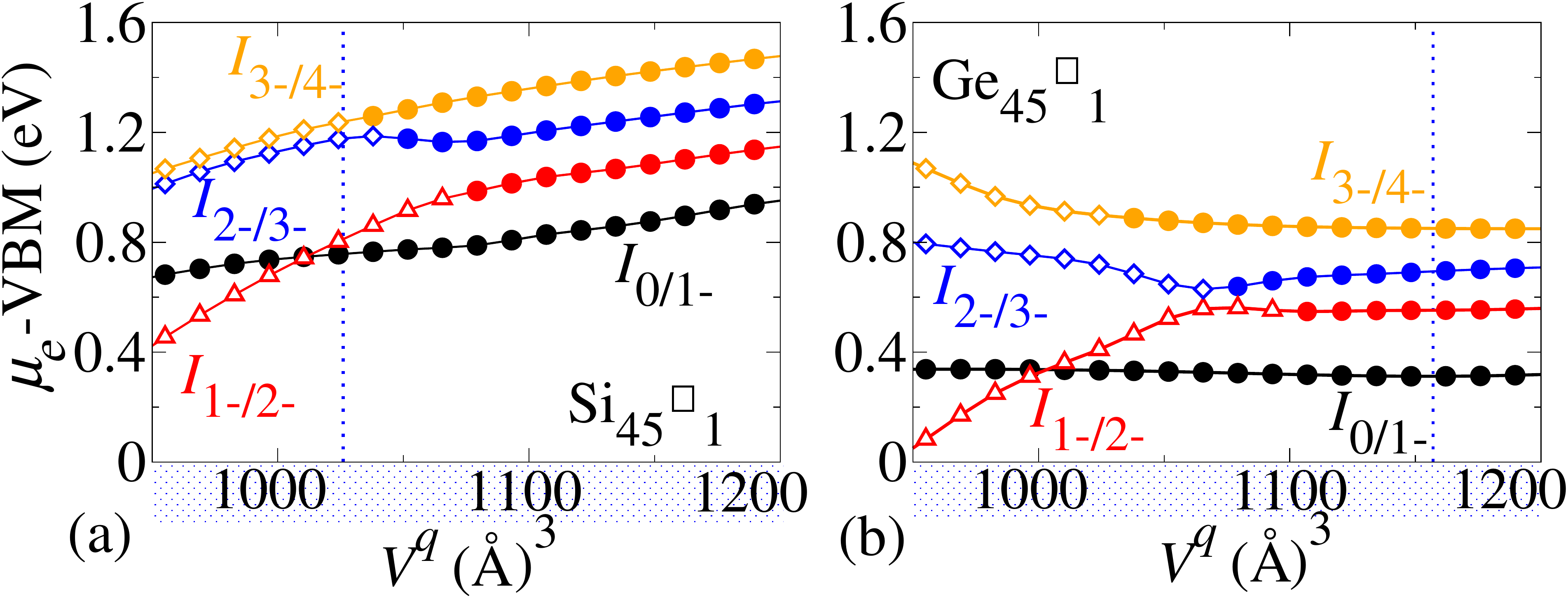}
\caption{\label{ctl_strained} Calculated~(DFT-PBEsol) charge transition levels $I_{q/q'}$ of (a)~$\mathrm{Si}_{45}\square_1$ and (b)~$\mathrm{Ge}_{45}\square_1$ as a function of unit cell volume $V^q$ with vacancies in charge state~$q$~($\bullet$ corresponds to a $D_{2d}$ symmetry, $\diamond$ to $D_2$ and $\triangle$ denotes
 the situation when $q$ and $q'$ differ in symmetry). The vertical lines denote the scenarios at the equilibrium-volume.}
\end{figure}

This peculiar and contrasting relaxation behavior of vacancies in clathrates can be attributed to the fact that the $3sp^3$~valence orbitals in Si$_{46}$ are spatially more localized than the $4sp^3$~orbitals in Ge$_{46}$. Accordingly, the Si framework is more rigid, which results in almost twice as high phonon frequencies~(see Suppl.~Mat.). In turn, this allows local atomic and global lattice degrees of freedom to decouple and to break the symmetry independently. Thus, vacancy creation in empty Si$_{46}$ leads to pairs of energetically unfavorable, spatially localized
non-interacting states~($I_{0/1-} \approx I_{1-/2-}$ and $I_{2-/3-} \approx I_{3-/4-}$, Fig.~\ref{ion_level}).
Conversely, the charged vacancy in Ge$_{45}\square_1$ forms a 
bulk state that retains the global symmetry and spans the vacancy in spite
of the large nearest neighbor distances. Simultaneously, this lowers
the charge transition levels upon relaxation. This interpretation in terms of localization of valence orbitals is substantiated by the volume dependence of the charge transition levels shown in Fig.~\ref{ctl_strained}. For all charge
states~$q$, we have computed~$E_\square^q$ at different volumes in \textbf{both} the lattice configurations with a local $D_2$ and $D_{2d}$ symmetry,~i.e.,~for
the different $c/a$ ratios and tiltings~$\gamma$ discussed before and listed in the Suppl.~Mat. The energetically most favorable~$E_\square^q$ is used to determine the transition levels~$I_{q/q'}$. Expanding $\mathrm{Si}_{45}\square_1$ and thus enforcing an increased delocalization results in $D_{2d}$-type
charge transition levels. Conversely, lattice compression~(enforced localization) leads to $D_2$-type levels in $\mathrm{Ge}_{45}\square_1$~(Fig.~\ref{ctl_strained}).
This shows that the discussed effects, which are driven by global and local symmetry breaking, are inherently related to the   
relationship between nearest neighbor distance~(more than $\sim$10 \% larger in clathrates than in the respective diamond structures) 
and valence orbital localization.

In summary, this works shows that vacancy formation is energetically not favorable
in Si clathrates, neither for K nor for Ba filling. Conversely, two vacancies per unit cell are formed in Ge clathrates fully filled with K,
and three vacancies in the case of Ba. In turn, this determines the different electronic character of these compounds. Regardless of the guest, 
the decisive energetic contribution for this contrasting behavior 
does not stem from the guests-host interaction or the formation of the neutral vacancy, but from its charging. 
The occurring electron transfer processes are quantitatively and qualitatively different in Si and Ge,
since their different degree of valence orbital localization leads to contrasting structural relaxation
patterns~(global and local symmetry breaking). 
The stoichiometry, thermodynamic stability, and electronic structure of these materials is thus determined by this microscopic mechanism, 
which arises from the large ratio of nearest neighbor distance and valence orbital localization. 
Accordingly, our study suggests that this mechanism can be influential also in other materials with elongated bonds,~e.g.,~skutterudites~\cite{Guodong2015}, 
Heusler alloys~\cite{Offernes2007}, boroncarbides~\cite{Balakrishnarajan2007}, and perovskites~\cite{Johnsson2007}. Also
in these cases, vacancy formation and its influence on structural and electronic properties are lively topics of research~\cite{Meisner1998,Galanakis2014,Schmechel1999,Eichel2011}.

\begin{acknowledgments}
The authors thank Sergey Levchenko and Patrick Rinke for many fruitful discussions. The work was partially funded by Einstein foundation (project ETERNAL) and the European Union's Horizon 2020 research and innovation program under grant agreement no. 676580, The Novel Materials Discovery (NOMAD) Laboratory, a European Center of Excellence.
\end{acknowledgments}

\end{document}